\documentclass[12pt,reqno]{amsart}

\usepackage{amsmath,amssymb}
\usepackage{epsfig}

\begin{document}

\title[]{Chaotic Phenomenon in Nonlinear Gyrotropic Medium}

\date{\today}

\maketitle


\author{{\large A. Ugulava,L. Chotorlishvili$^1$, K. Nickoladze, G. Mchedlishvili} \bigskip \\
{\it  ~Tbilisi State University,Department of Physics,
 Chavchavadze av. 3, 0128 Tbilisi, Georgia.\\E-mail lchotor33@yahoo.com$^1$} }

\begin{abstract}
Nonlinear gyrotropic medium is a medium, whose natural optical
activity depends on the intensity of the incident light wave. The
Kuhn's model is used to study nonlinear gyrotropic medium with
great success. The Kuhn's model presents itself a model of
nonlinear coupled oscillators. This article is devoted to the
study of the Kuhn's nonlinear model. In the first paragraph of the
paper we study classical dynamics in case of weak as well as
strong nonlinearity. In case of week nonlinearity we have obtained
the analytical solutions, which are in good agreement with the
numerical solutions. In case of strong nonlinearity we have
determined the values of those parameters for which chaos is
formed in the system under study. The second paragraph of the
paper refers to the question of the Kuhn's model integrability. It
is shown, that at the certain values of the interaction potential
this model is exactly integrable and under certain conditions it is
reduced to so-called universal Hamiltonian. The third paragraph
of the paper is devoted to quantum-mechanical consideration. It
shows the possibility of stochastic absorption of external field
energy by nonlinear gyrotropic medium. The last forth paragraph of
the paper is devoted to generalization of the Kuhn's model for
infinite chain of interacting oscillators.

\end{abstract}

\section*{Introduction}

In the week light flux, that propagates through medium, polarization of electromagnetic field does not depend on the intensity of light and is defined unambiguously by the polarization of the field incident on the interface of vacuum and the matter under study. This relationship is changed radically in nonlinear optics, where the indices of refraction and absorption are the functions of the radiation intensity. This may lead us to polarization multi-stability and polarization chaos. When the intensity in the  ``input'' is changed adiabatically in the ``output'' polarization is not yet the simple function and consists of stable and unstable areas [1]. At certain values of medium and radiation parameters, when we have in the ``input'' stationary radiation, in the ``output'' polarization undergoes auto-oscillatory changes or polarization pseudo-chaotic changes in time may happen, which have continuous frequency spectrum [1]. Polarization instability, due to nonlinear gyrotropy, is a matter of great interest for nonlinear optics of biological objects and exiton spectroscopy in solid states. The dependence of natural optical activity of isotropic medium on the intensity of electromagnetic wave is the simplest example of manifestation of nonlinear gyrotropy [2].

Nonlinear gyrotropy manifests itself clearly in such substances that are characterized by strong non-locality of nonlinear response. The ratio of characteristic size of medium  to the wavelength   plays the role of the spatial dispersion representative that determines the scale of non-locality. Therefore [3],  the effects of spatial dispersion of nonlinearity and polarization instability connected with them may be important first of all in the cholesterol liquid crystals near the phase point of transition into the isotropic state  $(\delta /\lambda \sim 10^{-1})$ , in chiral biological macromolecules $(\delta /\lambda \sim 10^{-1})$ and in semiconductors near the absorption resonances of exiton and bi-exitons $(\delta /\lambda\sim 10^{-1})$ [1]. The idea was developed to explain the polarization instability, according to which nonlinear gyrotropic medium may be
described as the unity of nonlinear oscillators presented by the Kuhn's nonlinear model [1]. The author of the mentioned article could obtain the implreit material equations of gyrotropic medium,
in which cubic nonlinearity of molecular oscillators is taken into account and which permits the indistinct solutions for medium response. First Kuhn introduced one of the popular oscillator models for optically active medium. Its detailed description may be seen in [4]. And we concentrate on some details. The Kuhn's molecule is formed by two oscillators, which are connected to each other by elastic interaction and they oscillate along the mutually perpendicular directions. The oscillators are separated in space. The Kuhn's molecule has right-left asymmetry i.e. it is a chiral molecule. The response of medium consisting of the Kuhn's molecules is non-local agitation of one of the oscillators conducted to the second end of the molecule. The size of the
molecule determines the characteristic scale of spatial dispersion. In spite of simplicity the Kuhn's model is quite productive. For example, the ensemble of the Kuhn's molecules oriented randomly represents the simplest model of isotropic gyrotropic medium. If we assume that the distance between the oscillators is zero, or if we rule out the bond between oscillators, the ensemble of Kuhn's molecules randomly oriented represents the simplest model of isotropic non-gyrotropic medium. Generalization of the Kuhn's model on the problems of nonlinear polarization optics presents one of the important directions of macroscopic modeling. This kind of generalization requires taking into account the nonlinear elastic forces in molecules. Let us examine the variant of nonlinear oscillator model according to the article [4]. The Hamiltonian of the molecule consists of kinetic energy of both charges, potential energy of elastic deformation and energy of oscillators interaction taking into account nonlinear members. We restrict ourselves to the inharmonious of individual oscillator and to the case of cubic nonlinearity bond between oscillators:
$$U(u_1,u_2)=\frac{1}{2}\omega_o^2(u_1^2+u_2^2)+\frac{a}{4}
(u_1^2+u_2^4)+\frac{c}{2}u_1^2u_2^2+b(u_1u_2^3+u_1^3u_2^)$$ here
$\omega_o$ is a natural frequency of individual oscillator,
$\alpha$ determine there interaction, $a,b,c$ determines
anharmonicity of oscillations. \vskip+2cm

\centerline{\epsfxsize=7cm\epsffile{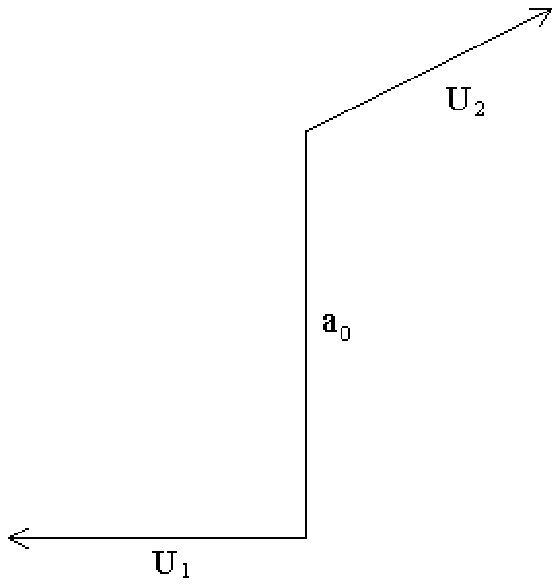}} \vskip+2cm

\centerline{Fig. 1 The diagram of the Kuhn coupled unlinear oscillator model}
\vskip+2mm

Function $U$ is invariant to the substitution of $u_1$ by $u_2$ and if the equilibrium positions $\vec{r}_1^{(0)}$ and $\vec{r}_2^{(0)}$   coincide with each other , molecule has  two planes of symmetry and the chirality does not  exist. But if $\vec{r}_1^{(0)}\not=\vec{r}_2^{(0)}$ and vectors $\vec{u}_1,\vec{u}_2$ and $\vec{r}_1^{(0)}-\vec{r}_2^{(0)}$ are noncollinear, the Kuhn's oscillators system loses symmetry with respect to any finite transformation and acquires spiral structure (chirality), i.e. it shall have different interaction with light of right and left polarization. In the simplest case the direction of the oscillations is orthogonal $u_1=\{x,0,0\}, u_2=\{0,y,0\}$ and equilibrium positions have coordinates $(0,0\pm a_o/2)$ ($a_o$- size of a molecule).

Having described briefly the Kuhn's model we remark that our aim is not the calculation of the optical parameters of medium. For us the Kuhn's model is a matter of direct interest as nonlinear
model, in which dynamical stochasticity may be formed. Our purpose is to determine the parameters of the problem for which:

1) system is precisely integrable;

2) chaos is formed in the system.

In spite of this success some questions are still open. The point is that, the Kuhn's generalized model [4] developed in [1] presents nonlinear dynamical system the study of which (not only
in the case of weak nonlinearity, when we can use the methods of perturbation theory, as was done in [1]), is a matter of interest itself. The Kuhn's nonlinear model realized in nonlinear optics
may be used for experimental observation of quantum chaos.

The subject of this paper is to study classical and quantum nonlinear dynamics in the limits of the Kuhn's generalized nonlinear model. The paper consists of the following paragraphs:

The first paragraph is devoted to study of classical nonlinear dynamics in case of weak nonlinearity, when we can consider the set of equations as quasi-linear. We have obtained, using Van-Der Pole generalized method [5, 6], the analytical solutions. In case of strong nonlinearity, when it is impossible to use analytical methods, numerical methods are used for this purpose. We have studied correlation functions of the solutions and we have determined the value of Kolmogorov-Sinai entropy [7,8]. Fractal dimension is determined Using Grassberger and Procaccia algorithm  [10--12].

In the second paragraph we have studied conditions of the Kuhn's model integrability. The exact analytical solutions written in the form of elliptical functions [13] are given. We have studied the
case, when amplitude of external field was modulated by slow changing radio wave field. We have studied  the possibility of formation of chaos in a one-dimensional nonautonomous system using
Melnikov's criteria [14]. It is shown that in case of the polychromatic pumping overlap of  the nonlinear resonances [15] occur and dynamical stochasticity was formed.

The third paragraph was devoted to quantum-mechanical study of our problem. In particular we have studied the case when amplitude of external field was modulated with slow variable field.

The last forth paragraph of the paper is devoted to generalization of the Kuhn's model for infinit clain  of interacting oscillators.

\section{The case of weak nonlinearity}

As we mentioned in introduction, the aim of this paper is to study the dynamics of the Kuhn's nonlinear oscillations influenced by the plane polarized light wave. Let us assume that the frequency of the external field is $\Omega$ then according to [1], the equations that describe the Kuhn's coupled nonlinear oscillations have the following form:
$$\ddot{x}+\omega_o^2x+\alpha y+\Gamma \dot{x}+ax^3+b(y^3+3x^2y)+cxy^2=\frac{eE_x}{m}\cos(\Omega
t+k_za_o/2),$$
\begin{equation}
\ddot{y}+\omega_o^2y+\alpha x+\Gamma
\dot{y}+ay^3+b(x^3+3y^2x)+cyx^2=\frac{eE_y}{m}\cos(\Omega
t-k_za_o/2),
\end{equation}
where $\omega_o$ is a natural frequency, $a,b,c$ are constants connected to anharmonic oscillations, $m$ and $e$ are mass and charge of electron, $E_{x,y}$ are the transverse components of the external electric field, $\Gamma$ is a phenomenological parameter, that describes dissipation, $k_z$ is wave vector projection, $a_o$ characteristic of spatial scale of the given problem. Let us consider that the following conditions are held:
 $$\frac{\Gamma}{\omega _o}\ll 1,~~\frac{eE_{x,y}}{m\omega_o^2}\ll 1.$$
Introducing dimensionless variables, with the help of the following transformation
$$x\rightarrow\frac{x}{a_o},~~y\rightarrow\frac{y}{a_o},~~t=\Omega t$$
equation (1) yields:
$$\ddot {x} + \frac{\omega_o^2}{\Omega^2}x + \frac{\alpha}{\Omega
^2}y=-f(x,y,\dot{x})+\frac{\omega^2}{\Omega ^2}\cos(t),$$
\begin{equation}
\ddot {y} + \frac{\omega_o^2}{\Omega ^2}y + \frac{\alpha}{\Omega
^2}x = -g(x,y,\dot{y})+\frac{\omega^2}{\Omega
^2}\frac{E_y}{E_x}\cos(t),
\end{equation}
where
$$f(x,y,\dot{x})=\frac{\Gamma}{\Omega}\dot{x}+A_ox^3+B_o(y^3+3x^2y)+C_oxy^2,$$
$$g(x,y,\dot{y})=\frac{\Gamma}{\Omega}\dot{y}+A_oy^3+B_o(x^3+3y^2x)+C_oyx^2,$$
$$ A_o\frac{aa_o^2}{\Omega^2},~B_o=b\frac{a_o^2}{\Omega^2},
~C_o=c\frac{a_o^2}{\Omega^2},~\omega=\biggl(\frac{eE_x}{a_om}\biggr)^{1/2}.$$
$f(x,y,\dot{x})$  and $g(x,y,\dot{y})$ are nonlinear functions of the coordinates. When the coefficient $\mu$ placed before these functions is equal to zero $\mu=0$, the set of equations (1)
becomes linear. When $\mu$ is not equal to zero but small, given set of equations is about linear one and that is why it is called quasi-linear. There are some methods of solution of quasi-linear
set of equations. For example asymptotic method, based on the hypothesis of existence such kind of solutions that are similar in form to the solutions of linear set of equations. There exists method that does not require existence of these kinds of solutions, but is based on finiteness of any real system's bandwidth (so-called filter hypothesis). We have used the method of slowly varying coefficients, which employs averaging approach. First this approach was used by Van-Der Pole for solving oscillation problems [5]. Let us rewrite the set of equations (2) in the generalized coordinates in the following form:
\begin{equation}
\left\{\begin{array}{l}
\ddot{q}_1\!+\!A_1\ddot{q}_2\!+\!B_1q_2\!+\!n_1^2q_1\!=\!\mu
f(q_1,\dot{q}_1,q_2,\dot{q}_2)\!+\!D_1\sin t \!+\!E_1\cos t, \\
\ddot{q}_2\!+\!A_2\ddot{q}_1\!+\!B_2q_1\!+\!n_2^2q_2\!=\!\mu
g(q_1,\dot{q}_1,q_2,\dot{q}_2)\!+\!D_2\sin t \!+\!E_2\cos t.
\end {array}\right.
\end{equation}
$$A_1=0,~ B_1=\frac{\alpha}{\Omega ^2},~
n_1^2=\frac{\omega_0^2}{\Omega^ 2},~ \mu=-1,~ D_1=0,~
E_1=\frac{\omega ^2}{\Omega ^2},~ q_1=x,$$ $$A_2=0,~
B_2=\frac{\alpha}{\Omega _2},~ n_2^2=\frac{\omega_0^2}{\Omega ^2},
~D_2=0;~ E_2=\frac{\omega ^2}{\Omega ^2}\frac {E_y}{E_x}=E_1\frac
{E_y}{E_x},~ q_2=y.$$ Let us look for solutions of the set (3)in
the form of solutions of linear set of equations: $$q_1=a\sin
(k_1t+\beta _1)+b \sin (k_2t +\beta _2) +d_1\sin t +e_1\cos t, $$
\begin{equation}
q_2=\alpha _1a\sin (k_1t+\beta _1)+ \alpha _2b \sin (k_2t +\beta
_2) +d_2\sin t +e_2 \cos t,
\end{equation}
where $a,b,\beta_1,\beta_2$ are slowly varying  in time functions, $k_1,k_2$ - the natural frequencies of linear uniform set of equations, which are solutions of the following equation: $$\sigma k^4-(n_1^2+n_2^2-A_1B_2-A_2B_1) k^2+n_1^2n_2^2-B_1B_2=0,$$
\begin{equation}
(\sigma=1-A_1A_2)
\end{equation}
and the rest of the coefficients are determined with the help of the following equations: $$\alpha
_1=\frac{A_2k_1^2-B_2}{n_2^2-k_1^2}=\frac{n_1^2-k_1^2}{A_1k_1^2-B_1},~~ \alpha
_2=\frac{A_2k_2^2-B_2}{n_2^2-k_2^2}=\frac{n_1^2-k_2^2}{A_1k_2^2-B_1},$$
$$d_1=\frac{D_1(n_2^2-1)-D_2(B_1-A_1)}{\Delta }, ~~d_2=\frac{D_2(n_1^2-1)-D_1(B_2-A_2)}{\Delta },$$
\begin{equation} e_1=\frac{E_1(n_2^2-1)-E_2(B_1-A_1)}{\Delta },~~e_2=\frac{E_2(n_1^2-1)-E_1(B_2-A_2)}{\Delta },
\end{equation}
$$\Delta = \sigma - (n_1^2+n_2^2-A_1B_2-A_2B_1)+n_1^2n_2^2-B_1B_2.$$ Substituting
these expressions in (5) yields: $$ \sigma = 1-A_1A_2=1, $$
 $$ k^4-\frac{2\omega _o^2}{\Omega  ^2}k^2+ \frac{\omega _o^4}{\Omega
^4}-\frac{\alpha ^2}{\Omega ^4}=0, $$ $$ k^2=\frac{\omega_o^2}{\Omega ^2}\pm\sqrt{\frac{\omega _o^4}{\Omega^4}-\frac{\omega _o^4}{\Omega ^4}+\frac{\alpha ^2}{\Omega ^4}}=\frac{\omega _o^2}{\Omega  ^2}\pm \frac{\alpha }{\Omega ^2}, $$
 $$k_1^2=\frac{\omega _o^2}{\Omega  ^2}+ \frac{\alpha }{\Omega ^2},~~
k_2^2=\frac{\omega _o^2}{\Omega  ^2}- \frac{\alpha }{\Omega ^2},$$  $$ \alpha _1= \frac{-B_2}{-\alpha/\Omega ^2}=1,~~ \alpha _2=\frac{-B_2}{\alpha/\Omega ^2}= -1, $$  $$ d_1=\frac{0}{\Delta }=0, ~~d_2=\frac{0}{\Delta }=0, $$ $$ \Delta =1-\frac{2\omega
_o^2}{\Omega ^2}+ \frac{\omega _o^4}{\Omega ^4}-\frac{\alpha
^2}{\Omega ^4}= \frac {(\Omega ^2-\omega _o^2)^2-\alpha ^2}{\Omega^4}, $$ $$ e_1=\frac{ \frac{\omega ^2}{\Omega ^2} (\frac{\omega_o^2}{\Omega ^2}-1)-E_1\frac{E_y}{E_x}\frac{\alpha}{\Omega^2}}{\Delta}=\frac{\omega ^2(\omega _o^2-\Omega  ^2-E^{*}\alpha)}{\Omega ^2-\omega _o^2-\alpha ^2}, $$ $$ e_2=\frac{
\frac{\omega ^2}{\Omega ^2}E^{*} (\frac{\omega_o ^2}{\Omega^2}-1)-\frac{\omega ^2}{\Omega ^2}\frac{\alpha}{\Omega^2}}{\Delta}=\frac{\omega ^2(E^{*}(\omega _o^2-\Omega  ^2)-\alpha)}{\Omega ^2-\omega _o^2-\alpha ^2},$$
 $$E^{*}=\frac{E_y}{E_x}.$$

As we have noted above, $a,b,\beta_1,\beta_2$ are slowly varying functions of time. Additional condition that will be imposed on these functions is the following: the time derivatives of the
generalized coordinates must have the same form as in the case of the constant $a,b,\beta_1,\beta_2$ coefficients. Taking into account the above mentioned, after time differentiation we obtain:
$$\dot{q}_1=ak_1\cos (k_1t+\beta _1)+bk_2\cos (k_2t+\beta
_2)-e_1\sin t, $$
$$\dot{q}_2=\alpha _1ak_1\cos (k_1t+\beta
_1)+\alpha _2bk_2\cos (k_2t+\beta _2)-e_2\sin t, $$
$$\dot{q}_1 =
\dot{a}\sin (k_1t+\beta _1)+a(k_1+\dot{\beta}_1)\cos (k_1t+\beta
_1)+\dot{b}\sin (k_2t+\beta _2) +$$
$$+ b(k_2+\dot{\beta}_2)\cos(k_2t+\beta _2) -e_1\sin t,$$
$$\dot{q}_2 = \dot{a} \alpha _1\sin
(k_1t+\beta _1)+a\alpha _1 (k_1+\dot{\beta}_1)\cos (k_1t+\beta
_1)+$$ $$+\dot{b}\alpha _2\sin (k_2t+\beta _2)
+ b\alpha
_2(k_2+\dot{\beta} _2)\cos (k_2t+\beta _2) -e_2\sin t,$$
from which we get the following condition:
$$\dot{a}\sin (k_1t+\beta
_1)+ a \dot{\beta}_1 \cos (k_1t+\beta _1)+ \dot{b}\sin (k_2t+\beta
_2)+ b \dot{\beta} _2 \cos (k_2t+\beta _2)=0,$$
$$ \dot{a}\alpha
_1\sin (k_1t+\beta _1)+ a\alpha _1 \dot{\beta}_1 \cos (k_1t+\beta
_1)+ \dot{b}\alpha _2\sin (k_2t+\beta _2)+$$
$$+ b \alpha _2\dot{\beta}_2 \cos (k_2t+\beta _2)=0. $$

Time differentiating of the time derivatives of the generalized coordinates $\dot{q}_1,\dot{q}_2$, substituting the obtained equations in the set (3), and taking into consideration (6) yields: $$\ddot{q}_1 = \dot{a}k_1\cos(k_1t+\beta_1)-
ak_1(k_1+\dot{\beta}_1)\sin(k_1t+\beta_1)+$$
$$\dot{b}k_2\cos(k_2t+\beta _2)-bk_2(k_2+\dot{\beta}_2)\sin
(k_2t+\beta_2)-e_1\cos t,$$ $$\ddot{q}_2=\dot{a}k_1 \alpha
_1\cos(k_1t+\beta_1)-ak_1\alpha
_1(k_1+\dot{\beta}_1)\sin(k_1t+\beta_1)+$$
\begin{equation}
\dot{b}k_2\alpha _2\cos (k_2t+\beta_2) - bk_2\alpha
_2(k_2+\dot{\beta}_2)\sin (k_2t+\beta_2)- e_2\cos t,
\end{equation}
$$\ddot{q}_1+B_1q_2+n_1^2q_1=\mu
f(q_1,\dot{q}_1,q_2,\dot{q}_2)+E_1\cos t,$$
$$\ddot{q}_2+B_2q_1+n_2^2q_1=\mu
g(q_1,\dot{q}_1,q_2,\dot{q}_2)+E_2\cos t,$$
$$\dot{a}k_1\cos(k_1t+\beta_1)+\dot{b}k_2\cos(k_2t+\beta_2) -
ak_1\dot{\beta}_1\sin(k_1t+\beta_1)-$$ $$- bk_2\dot{\beta}_2\sin
(k_2t+\beta_2)=\mu f(q_1,\dot{q}_1,q_2,\dot{q}_2),$$
$$\dot{a}\alpha _1 k_1\cos(k_1t+\beta_1)+\dot{b}k_2\alpha _2\cos
(k_2t+\beta_2)-ak_1\dot{\beta}_1\alpha _1\sin(k_1t+\beta_1)-$$
$$-bk_2\dot{\beta} _2\alpha _2\sin(k_2t+\beta_2)=\mu
g(q_1,\dot{q}_1,q_2,\dot{q}_2).$$ If we introduce new
variables $\xi=k_1t+\beta_1,~ \eta=k_2t+\beta_2$, then the set of
equations for $\dot{a},\dot{b},a\dot{\beta}_1,b\dot{\beta}_2$ is:
$$\dot{a}\sin\xi+a\dot{\beta}_1\cos \xi +\dot{b}\sin\eta +
b\dot{\beta}_2\cos\eta=0,$$ $$\dot{a}\alpha _1\sin\xi +a\alpha _1
\dot{\beta}_1\cos\xi +\dot{b}\alpha _2\sin\eta +b\alpha
_2\dot{\beta}_2\cos\eta=0,$$ $$\dot{a}k_1\cos\xi +\dot{b}k_2\cos
\eta -ak_1\dot{\beta}_1\sin\xi -bk_2\dot{\beta}_2\sin\eta =\mu
f(q_1,\dot{q}_1,q_2,\dot{q}_2),$$ $$\dot{a}\alpha _1 k_1\cos\xi +
\dot{b}k_2\alpha _2\cos\eta -ak_1\dot{\beta}_1\alpha _1\sin\xi
-bk_2\dot{\beta}_2\alpha _2\sin\eta =\mu
g(q_1,\dot{q}_1,q_2,\dot{q}_2),$$ The determinant of this set is:
$$\Delta =\left|\begin{array}{cccc}
\sin\xi&\sin\eta&\cos\xi&\cos\eta\\
\alpha_1\sin\xi&\alpha_2\sin\eta&\alpha_1\cos\xi&\alpha_2\cos\eta\\
k_1\cos\xi&k_2\cos\eta&-k_1\sin\xi&-k_2\sin\eta\\
\alpha_1k_1\cos\xi&\alpha_2k_2\cos\eta&-\alpha_1k_1\sin\xi&-\alpha_2k_2\sin\eta\\
\end{array}\right|=4k_1k_2.\,$$
$$\Delta_1=\left|\begin{array}{cccc} 0&\sin\eta&\cos\xi&\cos\eta\\
0&\alpha_2\sin\eta&\alpha_1\cos\xi&\alpha_2\cos\eta\\ \mu f
&k_2\cos\eta&-k_1\sin\xi&-k_2\sin\eta\\ \mu g
&\alpha_2k_2\cos\eta&-\alpha_1k_1\sin\xi&-\alpha_2k_2\sin\eta\\
\end{array}\right|=2k_2\cos\xi\mu(f+g)=$$
$$=-2k_2\cos(f+g),\,$$
$$\Delta_2=\left|\begin{array}{cccc} \sin\xi&0&\cos\xi&\cos\eta\\
\alpha_1\sin\xi&0&\alpha_1\cos\xi&\alpha_2\cos\eta\\
k_1\cos\xi&\mu f&-k_1\sin\xi&-k_2\sin\eta\\ \alpha_1k_1\cos\xi&\mu
g &-\alpha_1k_1\sin\xi&-\alpha_2k_2\sin\eta\\
\end{array}\right|=-2k_1\cos\eta\mu(g-f)=$$
$$=-2k_2\cos(f-g),\,$$
$$\Delta_3=\left|\begin{array}{cccc} \sin\xi&\sin\eta&0&\cos\eta\\
\alpha_1\sin\xi&\alpha_2\sin\eta&0&\alpha_2\cos\eta\\
k_1\cos\xi&k_2\cos\eta&\mu f&-k_2\sin\eta\\
\alpha_1k_1\cos\xi&\alpha_2k_2\cos\eta&\mu
g&-\alpha_2k_2\sin\eta\\
\end{array}\right|=2k_2\sin\xi\mu(f+g)=$$
$$=2k_2\sin\xi(f+g),\,$$
$$\Delta_4=\left|\begin{array}{cccc} \sin\xi&\sin\eta&\cos\xi&0\\
\alpha_1\sin\xi&\alpha_2\sin\eta&\alpha_1\cos\xi&0\\
k_1\cos\xi&k_2\cos\eta&-k_1\sin\xi&\mu f\\
\alpha_1k_1\cos\xi&\alpha_2k_2\cos\eta&-\alpha_1k_1\sin\xi&\mu g\\
\end{array}\right|=2k_1\sin\eta\mu(f-g)=$$
$$=2k_1\sin\eta(g-f),\,$$
and therefore $$\frac{da}{dt}=\frac{\Delta _1}{\Delta
}=\biggl[-\frac{1}{2k_1}(f+g)\biggr]\cos \xi,$$
$$\frac{db}{dt}=\frac{\Delta _2}{\Delta }=\biggl[
\frac{1}{2k_2}(f-g)\biggr]\cos \eta,$$
 $$ a\frac{d\beta
_1}{dt}=\frac{\Delta _3}{\Delta }=\biggl[
\frac{1}{2k_1}(f+g)\biggr]\sin \xi,$$ $$ b\frac{d\beta
_2}{dt}=\frac{\Delta  _4}{\Delta }=\biggl[-
\frac{1}{2k_2}(g-f)\biggr]\sin \eta.$$

The obtained equations are the set of equations (7) written in another variables. Suppose that $a,b,\beta_1,\beta_2$ the variables change slower in comparison with the oscillations taking
place in the initial dynamic system. If we average the obtained equations in periods $2\pi/k_1, 2\pi/k_2, 2\pi$, then we obtain so-called cut equations for $a,b,\beta_1,\beta_2$ variables:
$$\frac{da}{dt}=\biggl[-\biggl(\frac{1}{2k_1}\biggr)(F_1+G_1)\biggr],$$
\begin{equation}
\frac{db}{dt}=\biggl[\biggl(\frac{1}{2k_2}\biggr)(F_2-G_2)\biggr],
\end{equation}
$$\frac{d\beta_1}{dt}=\biggl[\biggl(\frac{1}{2k_1a}\biggr)(F_3+G_3)\biggr],$$
$$\frac{d\beta_2}{dt}=\biggl[\biggl(\frac{1}{2k_2b}\biggr)(G_4-F_4)\biggr],$$
where $$F_1=\frac{1}{4\pi ^3}\int \limits_o^{2\pi}\int
\limits_o^{2\pi}\int \limits_o^{2\pi} f\cos \xi d\xi d\eta dt,~~
F_2=\frac{1}{4\pi ^3}\int \limits_o^{2\pi}\int
\limits_o^{2\pi}\int \limits_o^{2\pi} f\cos \eta d\eta d\xi dt,$$
$$F_3=\frac{1}{4\pi ^3}\int \limits_o^{2\pi}\int
\limits_o^{2\pi}\int \limits_o^{2\pi} f\sin \xi d\xi d\eta dt,~~
F_4=\frac{1}{4\pi ^3}\int \limits_o^{2\pi}\int
\limits_o^{2\pi}\int \limits_o^{2\pi} f\sin \eta d\eta d\xi dt,$$
$$G_1=\frac{1}{4\pi ^3}\int \limits_o^{2\pi}\int
\limits_o^{2\pi}\int \limits_o^{2\pi} g\cos \xi d\xi d\eta dt,~~
G_2=\frac{1}{4\pi ^3}\int \limits_o^{2\pi}\int
\limits_o^{2\pi}\int \limits_o^{2\pi} g\cos \eta d\eta d\xi dt,$$
$$G_3=\frac{1}{4\pi ^3}\int \limits_o^{2\pi}\int
\limits_o^{2\pi}\int \limits_o^{2\pi} g\sin \xi d\xi d\eta dt,~~
G_4=\frac{1}{4\pi ^3}\int \limits_o^{2\pi}\int
\limits_o^{2\pi}\int \limits_o^{2\pi} g\sin \eta d\eta d\xi dt,$$

Taking into account the values of $\alpha$ coefficient, let us write coordinates, there derivatives, $f$ and $g$ functions: $$q_1
= a\sin \xi + b\sin \eta + e_1\cos t,$$ $$q_2 = a\sin \xi - b\sin
\eta + e_2\cos t,$$ $$\dot{q}_1 = ak_1\cos \xi + bk_2\cos \eta -
e_1\sin t,$$ $$ \dot{q}_2 = ak_1\cos \xi - bk_2\cos \eta - e_2\sin
t,$$
$$f(q_1,q_2,\dot{q}_1,\dot{q}_2)=\frac{\Gamma}{\Omega}(ak_1\cos\xi+bk_2\cos
\eta-e_1\sin t)+A_o(a\sin\xi+b\sin\eta+e_1\cos t)^3+$$
$$+B_o(a\sin \xi - b\sin \eta + e_2\cos t)^3 + 3B_o (a\sin \xi +
b\sin \eta + e_1\cos t)^2(a\sin\xi-b\sin \eta+e_2\cos t)+$$
$$+C_o(a\sin\xi+b\sin \eta + e_1\cos t) (a\sin \xi - b\sin \eta +
e_2\cos t)^2,$$ $$g(q_1,q_2,\dot{q}_1,\dot{q}_2) = \frac
{\Gamma}{\Omega}( ak_1\cos \xi - bk_2\cos \eta - e_2\sin t) +A_o
(a\sin \xi - b\sin \eta + e_1\cos t)^3 +$$ $$+B_o(a\sin \xi +
b\sin \eta + e_1\cos t)^3 + 3B_o (a\sin \xi - b\sin \eta + e_2\cos
t)^2(a\sin \xi + b\sin \eta + e_1\cos t) +$$ $$+C_o(a\sin \xi -
b\sin \eta + e_2\cos t) (a\sin \xi + b\sin \eta + e_1\cos t)^2$$

Substituting these functions and integrating yields the following results for $F$ and $G$ functions:
$$F_1= \frac{a\Gamma
k_1}{\Omega},~~ F_2= \frac{b\Gamma k_2}{\Omega},~~ G_1=
\frac{a\Gamma k_1}{\Omega},~~ G_2= -\frac{b\Gamma k_1}{\Omega},$$
$$F_3=\frac{a}{4}[6A_ob^2-2b^2c+3a^2(A_o+4B_o+C_o)+6(A_o+B_o)e_1^2+4(3B_o+C_o)e_1e_2+(6B_o+C_o)e_2^2],$$
$$F_4=\frac{b}{4}[6A_oa^2-2a^2c+3b^2(A_o-4B_o+C_o)+6(A_o-B_o)e_1^2+4(3B_o-C_o)e_1e_2+(-6B_o+2C_o)e_2^2],$$
$$G_3=\frac{a}{4}[6A_ob^2-2b^2c+3a^2(A_o+4B_o+C_o)+6(A_o+B_o)e_1^2+4(3B_o+C_o)e_1e_2+(6B_o+C_o)e_2^2],$$
$$G_4=\frac{b}{4}[-6A_oa^2+2a^2c-3b^2(A_o-4B_o+C_o)+(6B_o-2C_o)e_1^2+4(-3B_o+C_o)e_1e_2+(6B_o-2C_o)e_2^2].$$

Substituting the values of these functions in the set of equations (8)? yields:
\begin{equation}
\frac{da}{dt}= - \frac{a\Gamma}{\Omega},~~ \frac{db}{dt}=-
\frac{b\Gamma}{\Omega},
\end{equation}
$$\frac{d\beta
_1}{dt}=\frac{1}{8k_1}[6A_ob^2-2b^2C_o+3a^2(A_o+4B_o+C_o) +4
(3B_o+C_o) e_1e_2+(A_o+6B_o+C_o)(e_1^2+e_2^2)],$$ $$\frac{d\beta
_2}{dt}=\frac{1}{8k_2}[6A_oa^2+2a^2C_o-3b^2(A_o-4B_o+C_o) -4
(3B_o-C_o) e_1e_2+(6B_o-C_o-3A_o)(e_1^2+e_2^2)].$$

Separation of variables and integration of (9) yields:
\begin{equation}
a=e^{-\frac{\Gamma}{\Omega}t},~~b=e^{-\frac{\Gamma}{\Omega}t},
\end{equation}
$$\beta_1=\frac{1}{8k_1}[(-\frac{\Omega}{2\Gamma})\{6A_ob^2-2b^2C_o+3a^2(A_o+4B_o+C_o)\}+4(3B_o+C_o)e_1e_2+$$
$$+(A_o+6B_o+C_o)(e_1^2+e_2^2)]t,$$
$$\beta_2=\frac{1}{8k_2}[(-\frac{\Omega}{2\Gamma})\{6A_oa^2+2a^2C_o-3b^2(A_o-4B_o+C_o)\}-4(3B_o-C_o)e_1e_2+$$
$$+(6B_o-C_o-3A_o)(e_1^2+e_2^2)]t.$$

As we can see from the results obtained, in the time interval $0<t\ll\frac{\Omega}{\Gamma}$ it is possible to neglect relaxation processes, and effects of nonlinearity influence only the phases.
In the time intervals greater then the previous one $t\geq\frac{\Omega}{\Gamma}$ dissipation processes become significant. Due to decrease of the amplitude of the oscillation the influence of the nonlinear terms becomes insignificant. As a result, the forced oscillations of the linear coupled oscillators are obtained. This result obtained by qualitative analysis is in good agreement with the numerical solutions (see Fig. 2, Fig. 3).
\vskip+2cm

\centerline{\epsfxsize=10cm\epsffile{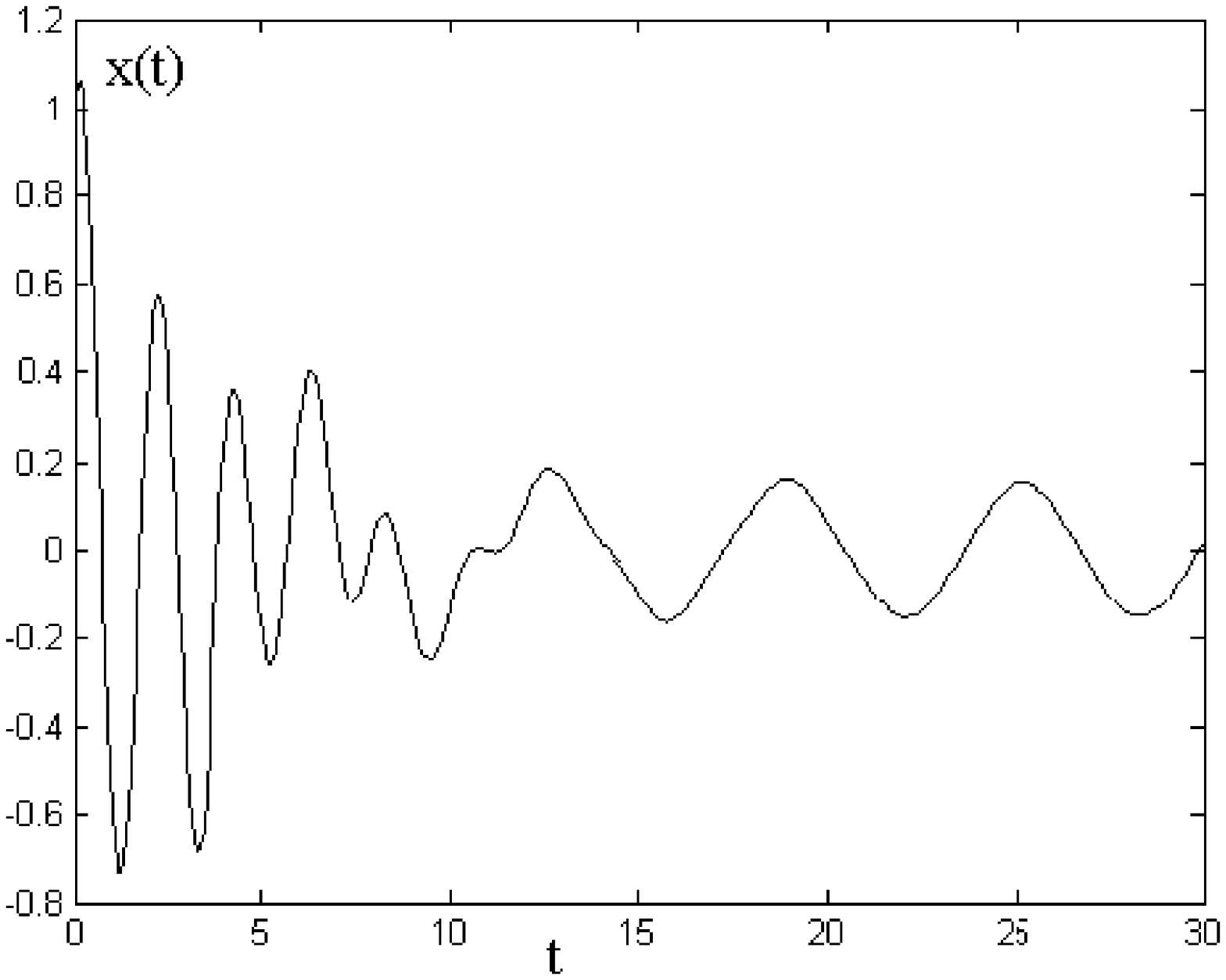}} \vskip+2cm

Fig. 2.  The plot of dependence of the oscillating system displacement (the nonlinear coupled oscillators) on the displacement's rate obtained by numerical integration of equation (1) for the following values of the parameters:
$$\frac{\omega_o^2}{\Omega ^2}=10,~~  \frac{\Gamma}{\Omega}=0.5,~~
\frac{\alpha}{\Omega ^2}=1,~~ \frac{\omega ^2}{\Omega ^2}=1.5,~~
E_y=2E_x,$$ $$ A_o,=0.3,~~ B_o=0.25,~~ C_o=0.25.$$
\vskip+2cm

\centerline{\epsfxsize=10cm\epsffile{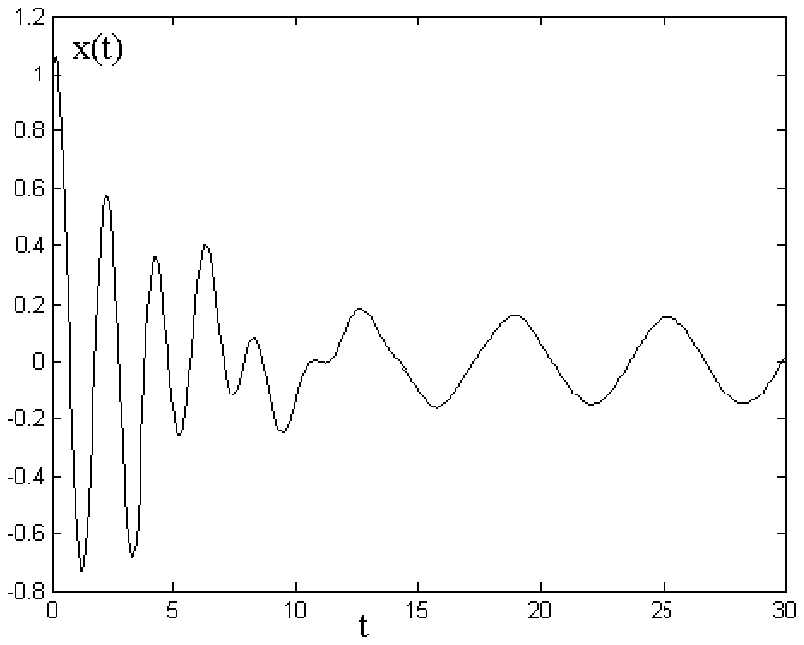}} \vskip+2cm

Fig. 3  The plot of dependence of the oscillating system displacement on time (the nonlinear coupled oscillators) obtained by numerical integration of equation (1) for the following values of the parameters:
$$\frac{\omega_o^2}{\Omega ^2}=10,~~
\frac{\Gamma}{\Omega}=0.5,~~ \frac{\alpha}{\Omega ^2}=1,~~
\frac{\omega ^2}{\Omega ^2}=1.5,~~  E_y=2E_x,$$ $$ A_o,=0.3,~~
B_o=0.25,~~ C_o=0.25.$$
\vskip+2cm

In case of strong nonlinearity it is impossible to use the method of slowly varying amplitudes and for description of system's behavior it is necessary to use numerical methods. In case of strong nonlinearity dynamical stochasticity may appear in the system. It is necessary to take stochastic description instead of dynamic one. For description of processes taken place in dynamical
system the following statistical quantities become important: Kolmogorov entropy [7], fractal dimension of strange attractor [10-12]. These quantities, statistical by definition, are closely
connected with the mechanical parameters of the system, specifically with local instability of phase trajectories. Our purpose is to determine those values of parameters of the given problem for which chaos is formed in the system described by the set of equations (1). Fig. 4 and Fig. 5 show these values obtained by numerical calculations.
\vskip+2cm

\centerline{\epsfxsize=10cm\epsffile{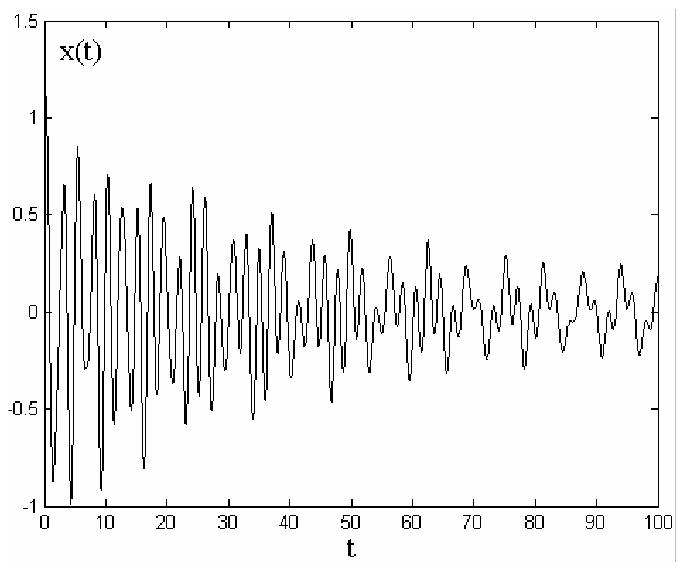}} \vskip+2cm

Fig. 4. The plot of dependence of the oscillating system displacement on time obtained by numerical integration of the set of equations (1) for the following values of parameters:
$$\frac{\omega_o^2}{\Omega ^2}=10,~~  \frac{\Gamma}{\Omega}=0.5,~~
\frac{\alpha}{\Omega ^2}=1,~~ \frac{\omega ^2}{\Omega ^2}=1.5,~~
E_y=2E_x,$$ $$ A_o,=3,~~ B_o=2.5,~~ C_o=2.5.$$
\vskip+2cm

\centerline{\epsfxsize=10cm\epsffile{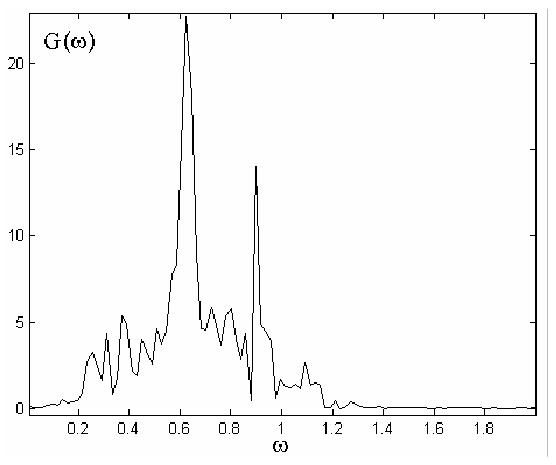}} \vskip+2cm

Fig. 5. Fourier image of correlation function obtained by integration of the set of equations (1) for the following values of the parameters. $$\frac{\omega_o^2}{\Omega ^2}=10,~~
\frac{\Gamma}{\Omega}=0.5,~~ \frac{\alpha}{\Omega ^2}=1,~~
\frac{\omega ^2}{\Omega ^2}=1.5,  E_y=2E_x,$$ $$ A_o,=3,~~
B_o=2.5,~~ C_o=2.5.$$
\vskip+2cm

In order to prove existence of chaos it is necessary to study correlation function of solutions. For the definition of Kolmogorov entropy we use Fourier transformation of correlation function:
\begin{equation}
G_x(\tau)=\langle x(t+\tau)x(t)\rangle,
\end{equation}
where $$\langle x(t+\tau)x(t)\rangle =\lim_{T\rightarrow\infty}\int\limits_o^T(x(t+\tau) x(t))dt,$$ is averaging in time $\tau_c$ is correlation time, which is connected with Kolmogorov
entropy through expression $h_o\sim\frac{1}{\tau}$. As numerical calculation shows correlation length is equal to $\tau_c=\frac{1}{\delta \omega}=0.83\times10^{-14}$ sec. The results of numerical calculations is shown on Fig. 6.

In order to calculate fractal dimension of the system's phase space we use Grassberger-Procaccia algorithm [10-12].

The sense of this algorithm is the following. Let's suppose we obtain state vectors ensemble  $\{x_i,~i=1,2,\ddot,N\}$ by numerical solving of the set of equations, that corresponds to
successive steps of integration of differential equation. Choosing small parameter $\varepsilon$ we can use our results for evaluation the following sum:
\begin{equation}
C(\varepsilon)=\lim_{N \rightarrow
\infty}\frac{1}{N(N-1)}\sum\limits _{i,j=1}^N
\theta(\varepsilon-|x_i-x_j|),
\end{equation}
where $\theta$ is a step function $\theta (x)=\left\{
\begin{array}{l} 0~~~x<0\\1~~~x\geq0 \end {array}
\right.$. According to Grassberger-Procaccia algorithm, if we know $C(\varepsilon)$ we can determine strange attractor's fractal dimension with the help of the following formula:
\begin{equation}
D=\lim_{\varepsilon \rightarrow
0}\frac{C(\varepsilon)}{log(\varepsilon)}.
\end{equation}
Let us calculate $C(\varepsilon)$ for different values of $\varepsilon$ and present the results in $log(\varepsilon)$ and $log(C(\varepsilon))$ coordinates. Supposed dependence
$C(\varepsilon)$ is a power function $\varepsilon^D$. So the obtained plot must be a line with slope $D$. The results of the numerical calculations are shown in Fig. 6.
\vskip+2cm

\centerline{\epsfxsize=10cm\epsffile{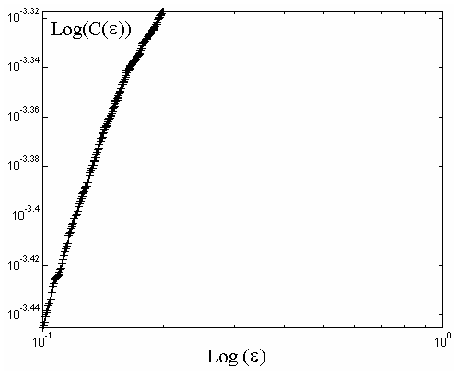}} \vskip+2cm

Fig. 6. The results of the numerical calculations in logarithmic scale. The results are obtained with numerical integration of the set of equations (1), using the formulae (12), (13) for the following values of the parameters.
$$\frac{\omega_o^2}{\Omega
^2}=10,~~ \frac{\Gamma}{\Omega}=0.5,~~ \frac{\alpha}{\Omega
^2}=1,~~ \frac{\omega ^2}{\Omega ^2}=1.5,~~  E_y=2E_x,$$ $$
A_o,=3,~~ B_o=2.5,~~ C_o=2.5.$$
\vskip+2cm

This plot enables us to calculate fractal dimension:
$$D=\frac{\Delta log(C(\varepsilon))}{\Delta
log(\varepsilon)}=2.4.$$

\section{ The condition of the Kuhn's model integrability}

Our next task is to determine the conditions of exact integrability of the set of equation (1). It is easy to see that in the absence of dissipation and external field, the Hamiltonian of the set of equations (1) has the following form:
\begin{equation}
H=\frac{\dot{x}^2}{2}+\frac{\dot{y}^2}{2}+\frac{\omega
_o^2}{2}(x^2+y^2) +axy + \frac{a} {4} (x^4+y^4)+b(xy^3+yx^3) +
\frac{c}{2}x^2y^2.
\end{equation}
Let us introduce new variables in order to separate them:
$$\left\{
\begin{array}{l}x=q_1+q_2\\y=q_1-q_2\end {array}
\right.$$ As a result (14) yields:
$$H=\dot{q}_1^2+\dot{q}_2^2+\alpha
q_1^2+\frac{aq_1^4}{2}+2bq_1^4+\frac{cq_1^4}{2}-\alpha
q_2^2+3aq_1^2q_2^2-cq_1^2q_2^2+$$
\begin{equation}
+\frac{aq_2^4}{2}-2bq_2^4+\frac{cq_2^4}{2}+
q_1^2\omega_o^2+q_2^2\omega_o^2.
\end{equation}
It is easy to see that when the following condition $c=3a$ is kept, our equation is integrable. When this condition is kept, the separated Hamiltonian is written in the following form:
\begin{equation}
H_1=\dot{q}_1^2+Bq_1^4+Aq_1^2=E_1,
~~H_2=\dot{q}_2^2+B^{*}q_2^4+A^{*}q_2^2=E_2,
\end{equation}
where $$B=\frac{a+4b+c}{2},~~A=\alpha  +\omega _o^2;$$
$$B^{*}=\frac{a-4b+c}{2},~~A^{*}=-\alpha  +\omega _o^2.$$

In consequence we have the following differential equation for the variable $q$. $$\frac{dq_1}{dt}=\sqrt{E_1-Bq_1^4-Aq_1^2}.$$ The
roots of the expression under root are:
$$(q_1^2)_{1,2}=\frac{A\pm\sqrt{A^2+4BE_1}}{-2B}$$

Taking into consideration this our differential equation has the following form: $$\frac{dq_1}{dt}=
\sqrt{(-B)\biggl(q_1^2-\frac{A+\sqrt{A^2+4BE_1}}{-2B}\biggr)\biggl(q_1^2-\frac{A-\sqrt{A^2+4BE_1}}{-2B}\biggr)}=$$
\begin{equation}
=\sqrt{B\biggl(\frac{A+\sqrt{A^2+4BE_1}}{2B}+
q_1^2\biggr)\biggl(\frac{\sqrt{A^2+4BE_1}-A}{2B}- q_1^2\biggr)}.
\end{equation}
Our aim is to find solutions of equation (16) in the form of elliptical function [13, 16]. Because of this let us introduce notations: $$\gamma_1= \frac{A+\sqrt{A^2+4BE_1}}{2B},~~ \gamma_2=
\frac{\sqrt{A^2+4BE_1}-A}{2B}.$$

Taking into account these notations equation (16) has the following form:
\begin{equation}
\frac{1}{\sqrt{B\gamma_1\gamma_2}}\frac{dq_1}{dt}=
\sqrt{\biggl(1+\frac{1}{\gamma_1}q_1^2\biggr)\biggl(1-\frac{1}{\gamma_2}q_1^2\biggr)}.
\end{equation}
It is readily seen that equation (17) can be brought to the following canonical form: $$\frac{dq_1}{dt}=\sqrt{(1-k^{\prime
2}q_1^2)(1+k^2q_1^2)},$$ where the coefficients $k$ and
$k^{\prime}$ must meet the condition: $$k^{\prime ^2}=1-k^2.$$ The
latter expression in our notation has the following form:
$$\frac{1}{\gamma_2}=1- \frac{1}{\gamma_1},$$
$$\frac{2B}{\sqrt{A^2+4BE_1}-A}=\frac{(A+\sqrt{A^2+4BE_1})-2B}{A+\sqrt{A^2+4BE_1}}.$$
This equation imposes the following condition on the value of energy: $$E_1=2B+\sqrt{4B^2+A^2}.$$ Let us evaluate the coefficient placed in front of time derivative of coordinate.
$$\frac{1}{\sqrt{B\gamma_1\gamma_2}}=\frac{1}{\sqrt{B
\frac{A+\sqrt{A^2+4BE_1}}{2B } \frac{\sqrt{A^2+4BE_1}-A}{2B}}} =
\frac{1}{\sqrt{E_1}}.$$

In consequence the equation has the form:
$$\frac{1}{\sqrt{E_1}}\frac{dq_1}{dt}=\sqrt{(1+k^2q_1^2)
(1-k^{\prime 2}q_1^2)},$$ here
$k^2=\frac{2B}{A+\sqrt{A^2+4BE_1}},~~ k^{\prime
2}=\frac{2B}{\sqrt{A^2+4BE_1}-A}.$

Analogous calculations for coordinate $q_2$ give: $$E_2=
\dot{q}_2^2+B^{*}q_2^4+A^{*}q_2^2,$$
$$\frac{dq_2}{dt}=\sqrt{E_2-Bq_2^4-Aq_2^2},$$
$$\frac{1}{\sqrt{E_2}}\frac{dq_2}{dt}=\sqrt{(1+k^{* 2}q_2^2)
(1-k^{\prime * 2}q_2^2)},$$ where $k^{* 2
}=\frac{2B^{*}}{A^{*}+\sqrt{A^{* 2}+4B^{*}E_2}},~~ k^{\prime * 2
}=\frac{2B^{*}}{\sqrt{A^{* 2}+4B^{*}E_2}-A^{*}}.$

Thus we obtain the following differential equations for $q_1$ and
$q_2$ coordinates: $$\frac{1}{\sqrt{E_1}}\frac{dq_1}{dt}=\sqrt{
(1+k^2q_1^2)(1-k^{\prime 2}q_1^2)},$$
$$\frac{1}{\sqrt{E_2}}\frac{dq_2}{dt}=\sqrt{(1+k^{* 2}q_2^2)
(1-k^{\prime * 2} q_2^2)}.$$

Separation of the variables and integration in the first equation gives: $$\int\frac{dq_1}{\sqrt{(1+k^2q_1^2)(1-k^{\prime 2}q_1^2)}}= \int\limits_o^t\sqrt{E_1}dt=\sqrt{E_1}t.$$ In consequence we obtain equation that determines the function $q_1(t)$ in the implicit form. Reversing this expression yields final solutions in
the form of elliptical function [14, 16]: $$q_1(t)=sd(k,\sqrt{E_1}t).$$ In its turn given Jacob elliptical function may be presented in the following form:
$$sd(k,t)=sn(k,t)/dn(k,t),~~ dn(k,t)= \sqrt{1-k^2sn^2(k,t)},$$
where $sn(k,t)$ is a Jacob sine function [14-16]. By analogy for the second coordinate we have: $$q_2(t)=sd(k^{*},\sqrt{E_2}t)=\frac{ sn(k^{*},\sqrt{E_2}t)}{ dn(k^{*},\sqrt{E_2}t)}.$$ Thus we have the following solutions: $$q_1(t)=sd(k,\sqrt{E_1}t)= \frac{sn(k,\sqrt{E_1}t)}{ dn(k,\sqrt{E_1}t)},$$
\begin{equation}
q_2(t)=sd(k^{*},\sqrt{E_2}t)= \frac{ sn(k^{*},\sqrt{E_2}t)}{
dn(k^{*},\sqrt{E_2}t)}.
\end{equation}
Based on these solutions let us evaluate whether the neglect of dissipation term is correct or not. Take into consideration that the period of the elliptical functions $dn(k,\sqrt{E}t)$ and
$sn(k,\sqrt{E}t)$ is $4K(k)$, where $K(k)$ is full elliptical integral [13, 14]. Compare this period $T=4K(k)$ to characteristic scale of time for which dissipation processes are considerable.
Taking into consideration equation (9) dissipation can be neglected if the following inequality is met: $4K(k)\gg\frac{\Omega}{\Gamma}$.

It can be seen readily that our problem is integrable if certain conditions are met. What will be consequences of taking into consideration external field and dissipate term, will chaos appear
in the system? We can make this question clear with the help of Melnikov's criterion [15, 16]. This criterion can be used to find out existence or absence of homoclinic structure, which is one of the interesting objects of nonlinear dynamics. Homoclinic structure can exist in phase space of dissipate as well as conservative systems. Existence of homoclinic structure can be used as evidence of existence of chaotic dynamics in a system that in turn implies existence of continuum of trajectories of random behavior. Melnikov's criterion reads: suppose we have one-dimensional integrable Hamiltonian $H(x,p)=E$, where $x=X(t)$ and $p=P(t)$ are solutions of undisturbed equations $\dot{x}=Hp,~~\dot{p}=-Hx;$ $f$, $g$ are periodic functions of time consisting of external field and dissipation, $\varepsilon$ is a small parameter $(\dot{x}=H_P+\varepsilon
f(x,p,t),~\dot{p}=-H_g+\varepsilon g(x,p,t))$. If autonomy of the system is breaking chaos is formed in the system when the following function changes sign
\begin{equation}
\Delta (\theta)=
\int\limits_{-\infty}^{+\infty}[H_X(X,P)f(X,P,t)+H_P(X,P)g(X,P,t)]dt,
\end{equation}
where $\theta$ is an initial time displacement. Thus using formula (1), (19), (20) the question of formation of chaos in the system is brought to study the behavior of the following function:
$$\Delta (\theta)= \int\limits_{-\infty}^{+\infty}[(\frac{\omega
^2}{\Omega^2}\cos(t+\theta)-\frac{\Gamma}{\Omega}\dot{q}_1(t+\theta))\dot{q}_1(t+\theta)]dt=$$
$$=\int\limits_{-\infty}^{+\infty}[(\frac{\omega ^2}{\Omega
^2}\cos(t+\theta)-
\frac{\Gamma}{\Omega}\frac{cn(k,t+\theta)}{dn^2(k,t+\theta)})
\frac{cn(k,t+\theta)}{dn^2(k,t+\theta)}]dt=$$
\begin{equation}
=\frac{\omega ^2}{\Omega
^2}\int\limits_{-\infty}^{+\infty}\frac{cn(k,t+\theta)}{dn^2(k,t+\theta)}\cos(t+\theta)dt
-\frac{\Gamma}{\Omega}\int\limits_{-\infty}^{+\infty}\frac{cn^2(k,t+\theta)}{dn^4(k,t+\theta)}dt
\end{equation}

The plot of the function $\Delta (\theta)$ is given on Fig. 7.
\vskip+2cm

\centerline{\epsfxsize=10cm\epsffile{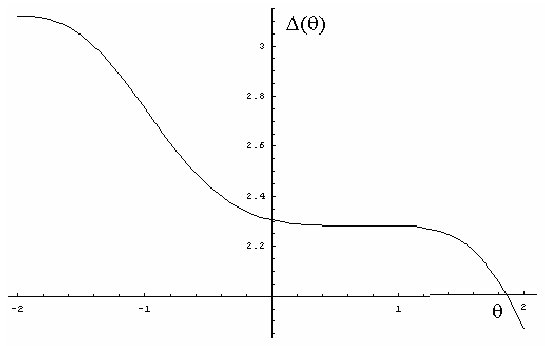}} \vskip+2cm

\begin{center}
Fig. 7.  Dependence of $\Delta (\theta)$ function on $\theta$ initial \\
time displacement.
\end{center}
\vskip+2mm

As can be seen from Fig. 7 this function $\Delta (\theta)$ does not change sign. This fact confirms that when there is certain symmetry in the system, chaos is not formed despite of strong nonlinearity and nonintegrability. Moreover, if the following condition is met $4K(k)\gg\frac{\Omega}{\Gamma}$ and it is possible to neglect dissipation, system is integrable even if it is affected by external field. In this case our Hamiltonian has the following form:
$$H(x,\dot{x},t)=H_o(x,\dot{x})+H_{NL}(x)+\varepsilon V(x,t),$$
where $$H_o=\frac{1}{2}(\dot{x}^2+Ax^2),~~H_{NL}=Bx^4;$$ $$
\varepsilon V(x,t)= \frac{\omega ^2}{\Omega ^2}x\cos(\Omega t),~~
\varepsilon=\frac{\omega ^2}{\Omega ^2};$$ $$ \varepsilon V(x,t)=
\varepsilon x\cos(\Omega t),~~\varepsilon\ll 1$$ $x,\dot{x}$ -
coordinate and momentum of electron, $\omega_o$ - natural frequency, $B$ - coefficient of nonlinear term, $\Omega$ -modulation frequency of variable field amplitude. We can consider
interaction of electron with external variable field as small perturbation. With the help of the canonical transformation of variables $x=(2I/\sqrt{A})^{1/2}\cos\theta,~~
\dot{x}=-(2I\sqrt{A})^{1/2}\sin\theta.$, we come to new variables action-angle $(I,\theta)$. In consequence we obtain: $H_o=I\sqrt{A}$. Considering that the resonance condition takes
place $\dot{\theta}\approx\Omega$, let us average the Hamiltonian  with the fast phase $\theta$ [17].$$ \overline{x^4} = \frac{1}{2\pi} \biggl(\frac{2I}{\sqrt{A}}\biggr)^2
\int\limits_o^{2\pi}\cos^4\theta d\theta = \biggl(\frac{2I}{\sqrt{A}}\biggr)^2\frac{3}{8}=
\frac{3}{2}\biggl(\frac{I}{\sqrt{A}}\biggr)^2,$$ $$
H_{NL}=\overline{ H_{NL}} =
\frac{3}{2}B\biggl(\frac{I}{\sqrt{A}}\biggr)^2.$$ Here averaging is equal to averaging in time $2\pi/\sqrt{A}=2\pi/\Omega$. This is small in this study, and $$\varepsilon V(x,t)=\varepsilon
\sqrt{\frac{2I}{A}}\cos\theta\cos(\Omega t)= \varepsilon
\sqrt{\frac{2I}{A}}\frac{1}{2}[\cos(\theta+\Omega t)+
\cos(\theta-\Omega t)]=$$ $$=\varepsilon V(I)[\cos(\theta+\Omega
t)+ \cos(\theta-\Omega t)],$$
$$H(I,\varphi,t)=H_o^{NL}+\varepsilon V(I)\cos\varphi,$$
$$H_o^{NL}=\sqrt{A}I+\overline{H}_{NL},~~
\overline{H}_{NL}=\frac{3}{2}(I/\sqrt{A})^2B,~~ V(I)=
(I/2\sqrt{A})^{1/2}.$$

Here we introduce a slowly varying phase $\varphi =\theta-\Omega t$, that changes slightly in time$\approx2\pi/\Omega.$ Because $\theta=\sqrt{A}t$ this corresponds to the resonance condition $\Omega=\sqrt{A}$.

Averaging with the fast phase $\theta$ corresponds to averaging in little time $2\pi/\sqrt{A}$. $\varphi$ changes a little at this time and that is why: $$\overline{\varepsilon V(x,t)}= \varepsilon
V(I,\varphi)=\varepsilon_o\sqrt{\frac{I}{2A}}\cos\varphi=\varepsilon
V(I)\cos\varphi,$$ $$
\overline{H(x,p,t)}=H(I,\varphi)=H_o^{NL}+\varepsilon
V(I,\varphi),$$ $$H_o^{NL}=\sqrt{A}I+\overline{H}_{NL}.$$ The
equations of motion have the following form: $$\left\{
\begin{array}{l}\dot{I}=-\varepsilon\frac{\partial}{\partial\varphi}V(I,\varphi),\\
\dot{\varphi}=\omega(I)+\varepsilon\frac{\partial}{\partial I
}V(I,\varphi).\end {array} \right.$$ The resonance condition is met for certain value $I_o$ that can be calculated from the same resonance condition: $\dot{\varphi}=\omega(I_o)$, where $$\omega (I)= \biggl(\frac{\partial H_o^{NL}}{\partial I}\biggr)-\Omega=
\sqrt{A}-\Omega +\frac{\overline{\partial  H_{NL}}}{\partial
I}=\sqrt{A}-\Omega+3B\frac{I}{A},$$ but resonance value of the action $I_o$ is calculated from the equality:$$\sqrt{A}-\Omega +3B\frac{I_o}{A}=0.$$ Deviation from the resonance action is $\Delta I=I-I_o, \Delta I\ll I_o$, only in this case $\varphi$ is small phase.

Taking into consideration the expression of $V(I,\varphi)$, the equations of motion have the form: $$\left\{ \begin{array}{l}\dot{I}=\varepsilon V(I)\sin\varphi,\\
\dot{\varphi}=\omega(I)+\varepsilon\frac{\partial V(I) }{\partial
I }\cos\varphi.\end {array} \right.$$

Expanding in a power series of $\frac{\Delta I}{I_o}\ll 1 $, and taking into account resonance condition $\omega(I_o)=0$, the set of equations takes the form: $$\left\{
\begin{array}{l}\Delta \dot{I}=\varepsilon V(I_o)\sin\varphi,\\ \dot{\varphi}=\omega^{\prime}\Delta I+\varepsilon\frac{\partial
V(I) }{\partial I }\biggr|_{I=I_o}\cos\varphi.
\end {array}
\right.$$

Let us evaluate approximations carried out. We have $V(I_o)\sim
H_o\sim I_o\sqrt{A},\frac{\Delta I_{max}}{I_o}\simeq
\frac{\sqrt{\varepsilon V(I_o)/\omega^{\prime}}}{I_o}\simeq
\sqrt{\frac{\varepsilon
I_o\sqrt{A}}{I_o^2\omega^{\prime}}}=\sqrt{\frac{\varepsilon}{\alpha}}\ll
1$.

Here we introduce dimensionless parameter of nonlinearity $\alpha \equiv (\frac{d\omega}{dI})\frac{I_o}{\sqrt{A}}$. Neglecting nonresonance terms can be made if the following condition is met for nonlinear resonance width $\Delta \omega= \sqrt{\varepsilon
V(I_o)\omega^{\prime}}\ll\sqrt{A}$ and relative width:
$$\frac{\Delta \omega}{\sqrt{A}}\simeq\frac{\sqrt{\varepsilon
V_o\omega^{\prime}}}{\sqrt{A}}=\sqrt{\frac{\varepsilon
I\sqrt{A}\omega^{\prime}}{A}}=\sqrt{\varepsilon\alpha} \ll 1$$.

Combining two inequalities $\sqrt{\frac{\varepsilon}{\alpha}}\ll 1$ and $\sqrt{\varepsilon\alpha} \ll 1$ we have $\sqrt{\varepsilon}\ll \sqrt{\alpha}\ll 1/\sqrt{\varepsilon}$ or $\varepsilon\ll\alpha\ll1/\varepsilon$. Taking into consideration the letter inequality, let us evaluate the second term in phase equation: $$\frac{\varepsilon \partial  V(I)/ \partial
I}{\omega^{\prime} \Delta I}\simeq \frac{\varepsilon V(I_o)/
I_o}{\omega^{\prime} \Delta I}\simeq \frac{\varepsilon
\sqrt{A}I_o/ I_o}{\omega^{\prime} \Delta I}= \frac{\varepsilon
\sqrt{A}I_o}{\omega^{\prime} \Delta I/I_o}=\simeq
\frac{\frac{\varepsilon}{\alpha}}
{\sqrt{\frac{\varepsilon}{\alpha}}}=
\sqrt{\frac{\varepsilon}{\alpha}}\ll 1.$$

Then the equations of motion have the form:
\begin{equation}\left\{
\begin{array}{l}\Delta \dot{I}=\varepsilon V(I_o)\sin\varphi\\
\dot{\varphi}=\omega^{\prime}\Delta I,
\end {array}
\right.\end{equation} and taking into consideration the results of this expansion, the Hamiltonian has the form:
\begin{equation}
H=\frac{\omega^{\prime}}{2}(\Delta I)^2+V\cos\varphi,
\end{equation}
where $V=\varepsilon V(I_o),
\omega^{\prime}=\frac{d\omega}{dI}|_{I=I_o}$. It can be readily seen that the given Hamiltonian, called universal Hamiltonian,  resembles the Hamiltonian of pendulum, which has mass $1/\omega^{\prime}$ and momentum $\Delta I$ and which is put in gravitational field with acceleration $g\sim V$. The quantity $1/\omega^{\prime}$ is measure of system's inertia: the more $1/\omega^{\prime}$, the more difficult the system to be taken out from the state of resonance with the help of pumping and vice versa. Taking into consideration the fact that in classical mechanics the problem of pendulum is exact solvable, our problem is brought to exact solvable problem. Coming out from this fact we can write solution of action directly:
\begin{equation}
\Delta I_+ = \sqrt{(E+\varepsilon V(I_o))/
\omega^{\prime}}dn(\sqrt{(E+\varepsilon V(I_o)) /
\omega^{\prime}t}; k)~~E>\varepsilon V(I_o)
\end{equation}
(with period $2K(1/k)$);
\begin{equation}
\Delta I_-=\sqrt{(E+\varepsilon V(I_o))/
\omega^{\prime}}cn(\sqrt{(E+\varepsilon V(I_o)) /
\omega^{\prime}t}; 1/k),~~E < \varepsilon V(I_o)
\end{equation}
(with period $4K(1/k)$), where $E$ is an energy, $cn$ and $dn$ are Jacoby elliptical functions (elliptical cosine and delta), $K(k)$ is full elliptical integral of the second order. When $E=\varepsilon V(I_o)$ or $k\rightarrow 1$ these two solutions are drawn near each other, and they have the form of instanton:
$$\Delta I_+ \rightarrow \Delta I_-\rightarrow\sqrt{2\varepsilon
V(I_o))/ \omega^{\prime}}/ ch(\sqrt{2\varepsilon V(I_o)) /
\omega^{\prime}t}).$$

\section{Quantum-mechanical consideration}

In the final part of previous paragraph we were able to show, that the study of the Kuhn's molecule influenced by external field is equivalent to that of the universal Hamiltonian when the potential of action has certain symmetry. The aim of this paragraph is quantum-mechanical study of the problem. After writing the stationary Schrodinger equation
\begin{equation}
\hat{H}\Psi_n=E_n\Psi_n
\end{equation}
for the universal Hamiltonian of the system Kuhn's molecule + pumping

\begin{equation}
\hat{H}=-\frac{\partial ^2}{\partial\varphi^2}+V,~~V=l\cos2\varphi
\end{equation}
we come to the equation coinciding with the  Mathieu's equation.
\begin{equation}
\frac{d^2\Psi_n}{d\varphi^2}+(E_n-V(l,\varphi))\Psi_n=0,
\end{equation}
where $E_n\rightarrow 8E_n/\hbar^2\omega^{\prime}$ is dimensionless quatity, $l$ is dimensionless amplitude of pumping, $\omega^{\prime}=\frac{d\omega(I)}{dI}$ is the derivative of nonlinear oscillation frequency $\omega (I)$ with respect to the action $I$. Characteristic peculiarity of the Mathieu-Schrodinger equation is the specific dependence of eigenvalues $E_n(l)$ and eigenfunctions $\Psi_n(\varphi,l)$ on parameter $l$ [18]. On the plane $(E,l)$, on which spectral indices are given, this peculiarity is manifested in the interchangeability of areas of
degenerated and nondegenerated states [19], Fig. 3, Fig. 4. The boundaries between these areas are passing through branching points of energetic terms $E_n(l)$. Let us suppose that pumping
amplitude is modulated with slow varying electromagnetic field. We can take into consideration influence of modulation if we make substitution in equation (28):
\begin{equation}
l(t)=l_o+\Delta l\cos\nu t,
\end{equation}
where $\Delta I$ is amplitude of modulation in dimensionless units, $\nu$- frequency of modulation. We imply, that little change $l(t)$ can encompass certain numbers of branching points $N$ on the right and the left side of the separatrix line (see [19], Fig. 3, Fig. 4.). Not going into details we only note that on $(E,l)$ plane different eigenfunctions correspond to different
areas. For particulars see [19]. Due to modulation of the parameter $l(t)$ the system goes from one area to another passing the points of branching. Thus in case of certain conditions:

1) amplitude modulated electric field acts on the system;

2) The coefficients of the potential of interaction satisfy the following requirements $c=3a$.

Then our problem is fully analogous with the problem studied in [20]. In the mentioned article, it was shown that, at the expense of indirect passage, energy of external field is absorbed and quantum chaos is formed. On the basis of aforesaid we can conclude, that experimental observation of quantum chaos is expected when amplitude modulated field acts on nonlinear
gyrotropic medium.

\section{Generalization of the Kuhn's model. The chain of interacting oscillators}

Our next task is to generalize the Kuhn's nonlinear model on the chain consisting $N$ oscillators. Our aim is to present nonlinear gyrotropic medium as ensemble of the nonlinear oscillators. The
question we are interested in is the following: how does initial perturbation propagate through such medium and whether system has soliton solutions or not? As an adiabatic approximation the
Hamiltonian consisting of $N$ oscillators has the form:
\begin{equation}
H=\sum_n(\frac{P_n^2}{2}+\frac{\omega
_o^2}{2}x_n^2+\frac{a}{4}x_n^4)+\sum_n[\alpha x_nx_{n+1} +
\frac{c}{2}x_n^2x_{n+1}^2+b (x_nx_{n+1}^3+ x_n^3x_{n+1})].
\end{equation}
We study the problem with fixed boundary conditions:
$$x_o=x_{N+1}=0,~~ \dot{x}_o=\dot{x}_{N+1}=0.$$

It is readily seen that the equations of motion of the Hamiltonian (30) have the form: $$ \ddot{x}_n=-[\omega
_o^2x_n+ax_n^3+\alpha(x_{n+1}+x_{n-1})+$$
\begin{equation}
+c(x_{n-1}^2x_n+x_nx_{n+1}^2)
+b(3x_{n-1}x_n^2+x_{n-1}^3)+b(x_{n+1}^3+3x_n^2x_{n+1})].
\end{equation}

In order to perform further calculations we introduce envelope function $\psi_i(t)=(-l)^ix_i(t)$ [21, 22], that is a function of coordinate $i$. We imply that envelope function is smooth function
of coordinate $i$. It enables us to proceed from discrete variable $\psi_i(t)$ to continuous one $x=li$, where $l$ is lattice period. Using Taylor expansion of continuous variable function in series we have:
\begin{gather}
\psi(x\pm l)= \psi (x)+\Psi (x)(\pm l)+\frac{1}{2}\psi _{xx}(x)
l^2+\notag\\
+\frac{1}{6}\psi_{xxx}(x)(\pm l)^3+ \dots
\end{gather}
Substituting formula (31) into (32), neglecting $l^4$ and terms of
higher power, we have:
\begin{equation}
\psi_{tt}+(\omega_o^2-2\alpha)\psi+(7a-8b)\psi^3-[\alpha+(6b-2c)\psi^2]
\psi_{xx}-(6b-2c)\psi\psi_x^2=0.
\end{equation}
With the aim to simplify our expression, let us suppose that the system has certain symmetry. In particular, let us assume that the coefficients of non-harmonic terms satisfy the following
requirement: $a=b=3c$. In order to solve equation (33) we use rotate wave approximation (RWA) [21,22]. In particular, let us find the solutions in the following form $\psi(x,t)=\psi(x)\cos\omega t$ and neglect the terms with multiple of $\omega$ arguments. As a result equation (33) yields:
\begin{equation}
(\omega_o^2-2\alpha
-\omega^2)\psi-\frac{3}{4}b\psi^3-\alpha\psi_{xx}=0.
\end{equation}
Let us show that equation (34) has exact analytical solutions that are periodic function of variable $x$. For this purpose we multiply equation (34) on $\psi_x$ and integrate the result:
\begin{equation}
\biggl(\frac{\omega^2+2\alpha
-\omega_o^2}{\alpha}\biggr)\psi^2+\frac{3b}{8\alpha}\psi^4+\psi_x^2=C_1.
\end{equation}
We are governed by article [22] to analyze further equation (35). Further we normalize $\psi_x$ function on its maximum value $\psi_m=\psi_{max}, f(x)=\frac{\psi(x)}{\psi_m}$. Then we rewrite this equation in the form that coincides with energy conservation law of point particle of unit mass put in external field:
\begin{equation}
H=\frac{f_x^2}{2}+U(f),
\end{equation}
where
\begin{equation}
U(f)=-\frac{3b}{16\alpha}\psi_m^2(1-f^2)(C_2+f^2)
\end{equation}
and
\begin{equation}
C_1=\biggl(\frac{\omega^2+2\alpha
-\omega_o^2}{\alpha}+\frac{3b}{8\alpha}\psi_m^2\biggr)\psi_m^2,
C_2=\frac{C_1}{\frac{3b}{8\alpha}\psi_m^2}
\end{equation}
solutions of equations (36) depend on the values of constant $C_2$.

It has been shown in [22] that when $C_2>0$, functions $\psi(x)$ are oscillating functions of $x$ which vary between minimum $\psi_m$ and maximum $-\psi_m$ values. The case $C_2=0$
corresponds to the separatrix solution, which is represented by the single localized breather with $\psi(x)\rightarrow 0$ at $x\rightarrow\pm \infty$. The third case
($C_2<0$) corresponds
to solutions that changes between nonzero values $\psi_{max}$ and $\psi_{min}$.

In order to satisfy boundary conditions (zero displacement in points $x=0$ and $x=N+1$), it is necessary to take the first case  ($C_2>0$) [22], because only in this case boundary conditions can be satisfied. Spatial period of this oscillation is given by this expression [22]
\begin{equation}
\int\limits_o^1df\biggl(\frac{df}{dx}\biggr)^{-1}=\Lambda /4.
\end{equation}
Zero boundary condition at point $x=N+1$ is met automatically, if half wavelength $\Lambda/2$ is the solution of the following equation: $(\Lambda/2 n)=N+1$, where $n=0,1,2 \dots N$.
Dissipation relation (39) determines spectrum of $\omega$ frequency as a function of $n$ and $\psi_m$.

If we express $f_x$ from equation (36) and substitute it into (39), after passing to new variable $\sin\alpha =f$, equation (39)  yields:
\begin{equation}
I(r,\psi_m)=\frac{2}{\pi}\int\limits_o^{\pi/2}\frac{d\alpha}{(\sin^2\alpha
+r^2
)^{1/2}}=\sqrt{\psi_m^2\frac{3b}{16\alpha}\biggl(\frac{N+1}{2n\pi}\biggr)^2}.
\end{equation}
Here we make substitution: $(\Lambda/4 (N+1)/2n$ and use notation $r^2=C_2$. Using the formulae (36), (40) we obtain the following expression for spatial profile function:
\begin{equation}
x(t)=\frac{1}{\sqrt{\frac{3b}{16\alpha}\psi_m^2}}
\int\limits_o^{arc\sin(\psi/\psi_m)}\frac{d\alpha}{(\sin^2\alpha
+r^2 )^{1/2}}.
\end{equation}
As we mentioned above, the solution of the breather type corresponds to condition $r^2=0$. This condition determines actually frequency of breather as a function of amplitude:
\begin{equation}
\omega_B^2=\omega_o^2-2\alpha -\frac{3b}{8}\psi_m^2.
\end{equation}
If we take into account the condition $r=0$ in expression (42), we obtain: $$x(f)=\frac{1}{\sqrt{\frac{3b}{16\alpha}\psi_m^2}}
\int\limits_{arc\sin f}^{\pi/2}\frac{d\alpha}{\sin\alpha},~~0\leq
x<\infty.$$ Integrating this expression we obtain finally:
\begin{equation}
\psi_B(x)=\psi_m\cos h^{-1}(\sqrt{\frac{3b}{16\alpha}\psi_m^2}x).
\end{equation}
Based on formulae (42), (43) we can conclude, that in non-gyrotropic medium excitations, whose frequency of propagation $\omega_B$ is determined by formula (42), during propagation
remains spatially localized.



\newpage

\begin{thebibliography}{99}

\bibitem{1} N. I. Zheludev, Usp. Fiz. Nauk 157, No.4, 683 (1989).

\bibitem{2} A. D. Petrenko and N. I. Zheludev, Optic. Acta. 31, 1174 (1984).

\bibitem{3} S. A. Akhmanov, N. I. Zheludev and R. S. Zadojan,
Zh. Eksp. Teor. Fiz., 91, 984 (1984).

\bibitem{4} Y. P. Svirko and N. I. Zheludev, Polarization of Light in
Nonlinear Optics (John Wiley Sons, Chichester, New York, Weinheim,
Brisbane, Singapure, Toronto) (1998).

\bibitem{5} N. V. Butenin, N. A. Fufaev, J. I. Neimark, Introduction to
the Oscillations Theory (Nauka, Moscow, 1987).

\bibitem{6} M. I. Rabinovich and D. I. Trubetzkov Introduction to
the Theory of Oscillations and Waves (Nauka, Moscow, 1984).

\bibitem{7} R. Z. Sagdeev, G. M. Zaslavsky, Introduction to the Nonlinear
Physics (Nauka, Moscow, 1988).

\bibitem{8} A. J. Lichtenberg, M. A. Lieberman, Regular and Stochastic Motion,
(Springer-Verlag, New York, Heidelberg, Berlin,1983).

\bibitem{9} H. G. Schuster, Determinictic Chaos (Physik-Verlag, Weinheim,
1984).

\bibitem{10} P. Grassberger, Phis. Lett. A97, 227 (1983).

\bibitem{11} P. Grassberger, Phis. Lett. A97, 224 (1983).

\bibitem{12} P. Grassberger and I. Procaccia, Phisica D, 9, 189 (1983).

\bibitem{13} N. Akhiezer, Elements of the Elliptical Functions Theory (Nauka, Moscow,
1970).

\bibitem{14} M. Abramovich and I. Stegun, Handbook of Mathematical
Functions (Washington, 1968).

\bibitem{15} S. P. Kuznetzov, Dynamical Chaos (Moscow, 2001).

\bibitem{16} J. I. Neimark and P. S. Landa, Stochastic and Chaotic
Oscillations (Nauka, Moscow, 1987).

\bibitem{17} G. M. Zaslavsky, Stochasticity of Dynamical Systems (Nauka, Moscow, 1984).

\bibitem{18} A. Ugulava, L. Chotorlishvili and K. Nickoladze, Phys. Rev.
E68, 026216 (2003).

\bibitem{19}  A. Ugulava, L. Chotorlishvili and K. Nickoladze, Phys. Rev.
E70,
026219 (2004).

\bibitem{20} A. Ugulava, L. Chotorlishvili and K. Nickoladze, Phys. Rev.
E71, 056211 (2005).

\bibitem{21} Y. A. Kosevich, Phys. Rev. B47, 3138 (1993).

\bibitem{22} A. J. Lichtenberg, V. V. Mirnov and C. Day, Chaos
15, 015109 (2005).
\end{thebibliography}
\end{document}